\documentclass[a4paper,11pt]{article}
\usepackage{graphicx,amssymb,amsmath,amsfonts,epsfig}
\usepackage[usenames,dvipsnames]{color}
\usepackage{subfigure}
\title{Area~(or entropy) products for Newman-Unti-Tamburino class of black holes }
\author{Parthapratim Pradhan\footnote{pppradhan77@gmail.com}\\ 
\textit{Department of Physics}\\
\textit{Hiralal Mazumdar Memorial College For Women}\\
{Dakshineswar, Kolkata-700035, India}}
\begin{document}

\maketitle

\begin{abstract}
We compute area~(or entropy) product formula for Newman-Unti-Tamburino~(NUT) 
class of black holes. Specifically, we derive the area product of outer horizon and
inner horizon~(${ \mathcal{H}}^{\pm }$) for Taub-NUT, Euclidean Taub-NUT black
hole, Reissner-Nordstr\"{o}m--Taub-NUT black hole, Kerr-Taub-NUT black hole
and Kerr-Newman-Taub-NUT black hole under the formalism developed
very recently by Wu et al.~\cite{wu}~[PRD
100, 101501(R) (2019)]. The formalism is that a generic four
dimensional Taub-NUT spacetime should be described completely in
terms of three or four different types of thermodynamic hairs. They
are defined as the Komar mass~($M=m$), the angular
momentum~($J_{n}=m\,n$), the gravitomagnetic charge ($N=n$), the
dual~(magnetic) mass $(\tilde{M}=n)$. After incorporating this
formalism, we show that the area~(or entropy) product of both the
horizons for NUT class of black holes are \emph{mass-independent}.
Consequently, the area product of ${\mathcal{H}}^{\pm }$ for these
black holes are \emph{universal}. Which was previously known in the
literature that the area product of said black holes are
\emph{mass-dependent}. Finally, we can say that this universality
is solely due to the presence of \emph{new conserved charges
$J_{N}=M\,N$} which is closely analogue to the Kerr like angular
momentum $J=a\,M$.
\end{abstract}

\vspace{2.5cm}
%PACS numbers: 04.70.Bw \hspace{0.3cm} 04.25.Nx \hspace{0.3cm} 04.40.Nr
\newpage

\textheight 25 cm

\clearpage

\section{Introduction}
Probably, black hole~(BH) is most intriguing as well as thermal objects~\cite{bcw73} 
in the universe. The thermal object by virtue of it has both temperature and 
entropy~\cite{bk73}. This entropy is again proportional to the 
 area of the event~(or outer) horizon~(EH) and Cauchy~(or inner) horizon~(CH) i.e. 
 ${\cal S}_{\pm} = \frac{{\cal A}_{\pm}}{4}$,
where, ${\cal S}_{\pm}$ is the Bekenstein-Hawking entropy~(in units in which $G=\hbar=c=k_{B}=1$) and 
${\cal A}_{\pm}$ is area of both the horizons. Analogously, the temperature is proportional to the 
surface gravity  of EH~(${\cal H}^{+}$) and CH~(${\cal H}^{-}$) i.e. 
$T_{\pm} = \frac{\kappa_{\pm}}{2 \pi}$, 
where $T_{\pm}$ is Hawking temperature evaluated on ${\cal H}^{\pm}$ and
$\kappa_{\pm}$ is surface gravity of the BH evaluated on ${\cal H}^{\pm}$.

Recently, it has been argued that a generic four dimensional Taub-NUT BH 
should be interpreted as in terms of 
three or four different types of thermodynamic hairs~\cite{wu}. They are defined as the Komar mass~($M=m$), 
the angular momentum~($J_{n}=m\,n$), the gravitomagnetic charge~($N=n$), the dual~(magnetic) 
mass~$(\tilde{M}=n)$. The authors also derived consistent thermodynamic first law  and Bekenstein-Smarr 
formula, Christodoulou-Ruffini mass formula and other thermodynamic properties. More appropriately,  
they solved long standing various problems of Taub-NUT~(both spherically symmetric and axisymmetric) geometry. 

In the Refs.~\cite{jetp,epl,mpla}, it has been demonstrated that the area~(or entropy) 
products for NUT 
class of BHs~(Taub-NUT,  Kerr-Taub-NUT and Kerr-Newman-Taub-NUT) have ``mass-dependent''
feature. This indicates that the area product of ${\cal H}^{\pm}$ has no nice 
universal feature nor any quantization feature. However in this work, we will 
prove that the area product of ${\cal H}^{\pm}$ for 
NUT class of BHs both in spherically symmetric cases and axisymmetric cases are 
indeed \emph{mass-independent}.  Therefore we can argue that the area product 
of ${\cal H}^{\pm}$ is  \emph{``universal''}  in nature. They depends only on the 
quantized charges and quantized angular momentum. Moreover, we argue that this 
universality occurs due to inclusion of \emph{spinning conserved charges $J_{n}$}. 
Alternatively, we can demand that the universality takes place due to  multihair attributes 
of the NUT parameter~(both rotation-like and electromagnetic charge like feature)~or
$$\frac{J_{n}}{m}=n\equiv N=\frac{J_{N}}{M}$$.

Area products of multi-horizon BHs are nowadays a fascinating topic of research for both general 
relativity community and string theory community due to the seminal work of Ansorg~\cite{ansorg09} 
and Cvetic~\cite{cvetic10}~[See also \cite{sd12,mv13,cr12,pope14,pp14,pp15,don,jetpl,abg,5d} ].  
The main focus therein is that the area~(or entropy) product of both the horizons
namely the EH~(${\cal H}^{+}$) and the CH~(${\cal H}^{-}$).
For instance, it is well-known that for 
Kerr-Newman class of BH having 
EH and CH, the area product formula of ${\cal H}^{\pm}$ becomes 
\begin{eqnarray}
{\cal A}_{+} {\cal A}_{-} &=& (8\pi)^2 \left[J^2+\frac{ Q^4}{4}\right] ~.\label{TNp}
\end{eqnarray}
The new physics here is that due to stationarity there is an absence of matter between 
the EH and CH~\cite{mv13}.

What is the origin of Bekenstein-Hawking~\cite{bk73,bcw73} entropy $S_{\pm}=\frac{A_{\pm}}{4}$ 
at the microscopic level? 
This is a genuine topic of research in quantum gravity. Here $A_{\pm}$ is the area of the 
BH EH and CH. There has been a significant success in four and 
five dimensions for 
supersymmetric Bogomol'ni-Prasad-Sommerfield~(BPS) class of  BHs~\cite{vafa}. Where 
it is demonstrated that the microscopic degrees of freedom should be explained
in terms of 2D conformal-field-theory~(CFT).
If there exists BH entropy~($S_{+}$) for outer horizon then it is also equally 
true that there exists BH entropy~($S_{-}$) for inner horizon. For maximally 
spinning spacetime, the microscopic entropy could be found in ~\cite{guica}. 
However, the detailed structure of the microscopic entropy of non-extremal BH still 
to be unknown to us. Some hopeful progress have been reported in ~\cite{castro10}. 

For  BPS class of BHs, the area product formula is computed to be~\cite{cvetic10}
\begin{eqnarray}
{\cal A}_{+} {\cal A}_{-}  &=& (8\pi)^2 \left[\sqrt{N_{1}}+\sqrt{N_{2}}\right]
\left[\sqrt{N_{1}}-\sqrt{N_{2}}\right]
= (8\pi)^2 N , \,\, N\in {\mathbb{N}}, N_{1}\in {\mathbb{N}}, N_{2} \in {\mathbb{N}} ~.\label{tnl}
\end{eqnarray}
Note that here the area product is synonymous to entropy product.  The integers $N_{1}$ and $N_{2}$ 
should be interpreted as the excitation numbers of 
the left and right moving modes of a weakly-coupled two-dimensional conformal field theory. 
The parameters $N_{1}$ and $N_{2}$ are depended explicitly on all the BH parameters.  This 
implies that the product must be quantized~\cite{finn97,cv96,cv97,cvf97} and it is an integer quantity.

However in this work, we will be focus on area product formula for NUT geometry. Particularly, we 
compute the product of all horizons areas for Taub-NUT BH, Euclidean Taub-NUT BH, 
Reissner-Nordstr\"{o}m--Taub-NUT BH, Kerr-Taub-NUT BH and  Kerr-Newman-Taub-NUT BH. 
Remarkably, our calculation shows that the area products for both the physical 
horizons are \emph{mass-independent~(universal)}. 
They depend only the various charges, angular momentum~($J$) and new 
conserved charges~($J_{n}$). This universal behaviour arises mainly due to the 
presence of new conserved charges $J_{N}$ which is closely analogue to the 
angular momentum $J=a\,M$ of Kerr class of BHs.

In the next section, we will compute 
the area product formula for Taub-NUT BH and show that the product is a universal quantity. 
Analogously, we derive the area products for  Euclidean-Taub-NUT 
BH in Sec.~(\ref{et}).
In Sec.~(\ref{ct}), we derive the area product formula for charged Taub-NUT BH and prove that 
the product is mass-independent. 
In Sec.~(\ref{kt}) and Sec.~(\ref{kntn}), we compute the area product 
formula for spinning Taub-NUT BH and charged  spinning Taub-NUT BH respectively. We prove that the 
product is indeed mass-independent for NUT class of BHs.  Finally, in Sec.~(\ref{con}), we have given 
the conclusions of the work.

\section{\label{ntn} Area product formula of Taub-NUT BH}
First, we consider the Lorentzian Taub-NUT BH~\cite{mis63,nut,mkg,ghz} 
\begin{eqnarray}
ds^2 &=& -{\cal B}(r) \, \left(dt+2n\,\cos\theta\, d\phi\right)^2+ 
\frac{dr^2}{{\cal B}(r)}+\left(r^2+n^2\right) \left(d\theta^2
+\sin^2\theta d\phi^2 \right) ~,\label{tn1}
\end{eqnarray}
where the function ${\cal B}(r)$ is defined by 
\begin{eqnarray}
 {\cal B}(r) &=& \frac{1}{r^2+n^2} \left[r^2-n^2-2mr \right]
\end{eqnarray}
Under new formalism~\cite{wu} for NUT class of BHs, the global 
conserved charges are defined as 
\begin{eqnarray}
m &=& M~\mbox{(Komar mass)} \nonumber\\
n &=& N ~\mbox{(gravitomagnetic charge)}\nonumber\\
m\,n &=& J_{n}~ \mbox{(angular momentum)} ~.\label{tn2}
\end{eqnarray}
Taking  cognizance of Eq.~({\ref{tn2}}) then the metric 
can be re-written as 
\begin{eqnarray}
ds^2 &=& -{\cal B}(r) \, \left(dt+2N\,\cos\theta d\phi\right)^2+ \frac{dr^2}{{\cal B}(r)}+\left(r^2+N^2\right)
\left(d\theta^2+\sin^2\theta d\phi^2 \right) ~,\label{tn3}
\end{eqnarray}
where the function ${\cal B}(r)$ is given by 
\begin{eqnarray}
{\cal B}(r) &=& \frac{1}{r^2+N^2} \left(r^2-N^2-2Mr\right)
\end{eqnarray}
The BH horizons are located at 
\begin{eqnarray}
r_{\pm}= M \pm \sqrt{M^2+N^2}\,\, \mbox{and}\,\,  r_{+}> r_{-}
\end{eqnarray}
$r_{+}$ is called EH  and $r_{-}$  is called CH. Thus the area of the BH is computed as 
\begin{eqnarray}
{\cal A}_{\pm} &=& \int^{2\pi}_0\int^\pi_0\sqrt{g_{\theta\theta}\,g_{\phi\phi}}{\mid }_{r=r_{\pm}} d\theta\, d\phi 
                = 8\pi\left[M^2+N^2 \pm \sqrt{M^4+J_{n}^2}\right]
~.\label{ctn}
\end{eqnarray}
Thus the area product of  ${\cal H}^{\pm}$ is computed to be 
\begin{eqnarray}
{\cal A}_{+}{\cal A}_{-} &=& (8\pi)^2\left(J_{n}^2+N^4\right)~. \label{ntn1}
\end{eqnarray}
It implies that the product is indeed independent of the Komar mass. Consequently, the product 
is  \emph{universal}. From the area product formula it follows that this expression is truly
universal solely due to
$$J_{n}=m\,n=M\,N=J_{N}$$
which is quite analogue to the Kerr like angular momentum. Alternatively, we can say 
that the universality comes due to the multihair attributes of NUT parameter.  
Except this new input the area product of NUT class of BHs never be universal.

The BH temperature of both the horizons are computed via surface gravity on 
the horizons  defined by  
\begin{equation}
T_{\pm}=\frac{\kappa_{\pm}}{ 2\pi}= \frac{{\cal B}'(r_{\pm})}{4\pi\left(r_{\pm}^2+N^2\right)}
=\frac{r_{\pm}-M}{2\pi\left(r_{\pm}^2+N^2\right)}
=\frac{1}{4\pi\,r_{\pm}},
\end{equation}
It should be noted that besides the area of ${\cal H}^{\pm}$, the surface gravity of ${\cal H}^{\pm}$ 
is also another important quantity of the thermodynamics. Therefore it may play crucial role to 
understanding the entropy at the microscopic level. 
Almost all the other thermodynamic properties like first law of thermodynamics, Smarr formula etc. 
have discussed  in Ref.~\cite{wu} so we have not repeated them here.

\section{\label{et} Area products for Euclidean Taub-NUT BH}
The Euclidean~\cite{hunter98,hunter99,page99,myers99,empa99} class of Taub-NUT BH should be 
obtain by putting the condition
$t=i\,\tau$ and $N=i\,N$ in Eq.~(\ref{tn3}). Then 
the function ${\cal B}(r)$ becomes
\begin{eqnarray}
 {\cal B}(r) &=& \frac{1}{r^2-N^2} \left(r^2+N^2-2Mr\right)
\end{eqnarray}
Therefore  the BH horizons are situated at 
\begin{eqnarray}
r_{\pm}=  M\pm\sqrt{M^2-N^2}\,\, \mbox{and}\,\,  r_{+}> r_{-}
\end{eqnarray}
Where $r_{+}$ is called EH  and $r_{-}$  is called CH. Thus the area of the BH is computed as 
\begin{eqnarray}
{\cal A}_{\pm} &=& \int^{2\pi}_0\int^\pi_0\sqrt{g_{\theta\theta}\,g_{\phi\phi}}{\mid }_{r=r_{\pm}} d\theta\, d\phi 
                = 8\pi\left[M^2-N^2 \pm \sqrt{M^4-J_{N}^2}\right]
~.\label{ctne}
\end{eqnarray}

Therefore the area product of  ${\cal H}^{\pm}$ becomes 
\begin{eqnarray}
{\cal A}_{+}{\cal A}_{-} &=& (8\pi)^2\left(N^4-J_{n}^2\right)~. \label{ntn2}
\end{eqnarray}
Thus it implies that the product is mass-independent and consequently it is
universal.

The BH temperature is defined by  
\begin{equation}
T_{\pm}=\frac{\kappa_{\pm}}{ 2\pi}= \frac{{\cal B}'(r_{\pm})}{4\pi\left(r_{\pm}^2-N^2\right)}
=\frac{r_{\pm}-M}{2\pi\left(r_{\pm}^2-N^2\right)}=\frac{1}{4\pi\,r_{\pm}},
\end{equation}

\section{\label{ct} Area product formula of  Reissner-Nordstr\"{o}m--Taub-NUT BH}
Now we can extend the area product formula to the charged case. Therefore the function ${\cal B}(r)$ 
in this case would be
\begin{eqnarray}
 {\cal B}(r) &=& \frac{1}{r^2+N^2} \left(r^2-N^2-2Mr+Q^2 \right)
\end{eqnarray}
where $Q$ is the purely electric charge. Now the BH horizons are located at 
\begin{eqnarray}
r_{\pm}= M \pm\sqrt{ M^2-Q^2+N^2}\,\, \mbox{and}\,\,  r_{+}> r_{-}
\end{eqnarray}
Now the area of the BH is 
\begin{eqnarray}
{\cal A}_{\pm} &=& \int^{2\pi}_0\int^\pi_0\sqrt{g_{\theta\theta}\,g_{\phi\phi}}{\mid }_{r=r_{\pm}} d\theta\, d\phi 
                = 4\pi\left[2 \left(M^2+N^2 \right)-Q^2 \pm 2\sqrt{M^4-M^2Q^2+J_{N}^2}\right] ~.\label{tnRN}
\end{eqnarray}
Hence the area product of  ${\cal H}^{\pm}$ is 
\begin{eqnarray}
{\cal A}_{+}{\cal A}_{-} &=& (8\pi)^2\left[J_{n}^2+\left(\frac{Q^2}{2}-N^2\right)^2\right]
~.\label{tnrn}
\end{eqnarray}
Therefore the product of ${\cal H}^{\pm}$ is mass-independent as well as universal.

\section{\label{kt} Area product formula of Kerr-Taub-NUT BH}
Now we turn to the spinning NUT BH. The metric for spinning Taub-NUT BH~\cite{miller} in
Boyer-Lindquist like coordinates $(t, r, \theta, \phi)$ is 
\begin{eqnarray}
ds^2 &=& -\frac{\Delta}{\rho^2} \, \left[dt-P d\phi \right]^2+\frac{\sin^2\theta}{\rho^2}\,
\left[(r^2+a^2+N^2)\,d\phi-a\,dt\right]^2+\rho^2 \, \left[\frac{dr^2}{\Delta}+d\theta^2\right]
~.\label{ktn}
\end{eqnarray}
where
\begin{eqnarray}
a &\equiv&\frac{J}{M},\, \rho^2 \equiv r^2+(N+a\,\cos\theta)^2 \\
\Delta &\equiv& r^2-2Mr+a^2-N^2\\
P & \equiv& a\,\sin^2\theta-2N\,\cos\theta  ~.\label{ad}
\end{eqnarray}
where the global conserved charges for this spacetime are the Komar mass $M$, angular momentum $J=a M$ and 
gravitomagnetic charge or dual mass or NUT parameter $N$. 

The radii of the horizon is computed by the function
$$\Delta|_{r=r_{\pm}}=0$$
which implies that
\begin{eqnarray}
r_{\pm} &\equiv&  M \pm\sqrt{ M^2-a^2+N^2}
\end{eqnarray}
Now after introducing the new conserved charge $J_{N}=MN$,  the area of both the horizons~(${\cal H}^\pm$) is
\begin{eqnarray}
{\cal A}_{\pm} &=& \int^{2\pi}_0\int^\pi_0 \sqrt{g_{\theta\theta}\,g_{\phi\phi}}\, d\theta\, d\phi  \\
                &=& 8\pi\left[(M^2+N^2) \pm  \sqrt{M^4+J_{N}^2-J^2} \right] ~.\label{arKTN}
\end{eqnarray}
Now we can calculate the area product for Kerr-Taub-NUT BH as
\begin{eqnarray}
{\cal A}_{+} {\cal A}_{-} &=& (8\pi)^2 \left(J^2+J_{N}^2+N^4\right) ~.\label{pKtn}
\end{eqnarray}
The area product is mass-independent as well as universal.

\section{\label{kntn} Area product formula of Kerr-Newman-Taub-NUT BH}
Finally, we consider most general class of BH without introducing the cosmological constant. The metric is of the 
form like Eq.~(\ref{ktn}) and the horizon function~\cite{miller} is defined to be 
\begin{eqnarray}
\Delta &\equiv&  r^2-2 M r+a^2+Q^2-N^2  ~.\label{delknn}
\end{eqnarray}
Here the conserved charges are Komar mass $M$, angular momentum $J=a M$ and 
gravitomagnetic charge or dual (magnetic) mass or NUT charge $N$. The horizons 
are located at 
\begin{eqnarray}
r_{\pm} &\equiv &  M \pm \sqrt{M^2-a^2-Q^2+N^2}
\end{eqnarray}
Analogously, the BH horizon area of ${\cal H}^\pm$ is derived to be 
\begin{eqnarray}
{\cal A}_{\pm} &=& \int^{2\pi}_0\int^\pi_0 \sqrt{g_{\theta\theta}\,g_{\phi\phi}} d\theta d\phi \\
                &=& 4\pi\left[2(M^2+N^2)-Q^2 \pm 2 \sqrt{M^4+J_{N}^2-J^2-M^2Q^2} \right]
~.\label{arKNTN}
\end{eqnarray}
Finally, the product of area of both the horizons for most general class of NUT BH without 
the cosmological constant
\begin{eqnarray}
{\cal A}_{+} {\cal A}_{-} &=& (8\pi)^2\left[J^2+J_{N}^2+\left(\frac{Q^2}{2}-N^2\right)^2\right] 
~.\label{prKNtn}
\end{eqnarray}
The product is universal as well as quantized.

\section{\label{con} Conclusions}
It has recently been stated that a generic 4D Taub-NUT BH could be completely
described in terms of three or four different types of thermodynamic hairs. They must be defined 
as the Komar mass~($M=m$), the angular momentum~($J_{n}=m\,n$), the gravitomagnetic charge ($N=n$), 
the dual~(magnetic) mass $(\tilde{M}=n)$. Under these circumstances, we 
demonstrated that the area ~(or entropy) product 
of multi-horizons for NUT class of BH is \emph{mass-independent}. 
That means the mass-independent features of NUT class of BHs are interestingly 
shown to be \emph{universal} which was earlier known as ``mass-dependent'' 
characteristics in the literature. 

We have explicitly checked here for Lorentzian Taub-NUT BH, Euclidean Taub-NUT BH, Reissner-Nordstr\"{o}m--Taub-NUT BH, 
Kerr-Taub-NUT BH and  Kerr-Newman-Taub-NUT BH. In each case, we showed that the area product of 
both physical horizons is universal in nature after the inclusion of angular momentum like parameter 
$J_{N}=M\,N$ with the generic parameters $M$ and $N$ as a global conserved charges.  This universal feature 
of area ~(or entropy) products that we have derived for NUT class of BHs further provide any hints 
for potential explanation of the microscopic structure of the BHs in four dimensions.

\end{document}